\documentclass{ws-ijmpaMY}
\usepackage{hyperref}

\begin{document}
\markboth{A. A. Grib \& Yu. V. Pavlov}{Particle properties outside of
the static limit in cosmology}

% \catchline{}{}{}{}{}
%
%%%%%%%%%%%%%%%%%%%%%%%%%%%%%%%%%%%%%%%%%%%%%%%%%%%%%%%%%%%%%%%%%%%%

\title{Particle properties outside of the static limit in cosmology}

\author{A. A. Grib}

\address{A.~Friedmann Laboratory for Theoretical Physics, Saint Petersburg, Russia;\\
Theoretical Physics and Astronomy Department of the Herzen  University,\\
Moika 48, Saint Petersburg 191186, Russia\\
andrei\_grib@mail.ru}

\author{Yu. V. Pavlov}

\address{Institute of Problems in Mechanical Engineering, Russian Academy of Sciences,\\
Bol'shoy pr. 61, St. Petersburg 199178, Russia;\\
N.I.\,Lobachevsky Institute of Mathematics and Mechanics,
    Kazan Federal University,\\
    18 Kremlyovskaya St., Kazan 420008, Russia\\
yuri.pavlov@mail.ru}

\maketitle

\begin{history}
\received{Day Month Year}
\revised{Day Month Year}
\end{history}

\begin{abstract}
    It is shown that in the rest frame of the observer in expanding Universe
states of particles with negative energy exist.The properties of such states
are studied.
    The comparison  with the case of negative energies of particles in black
holes and rotating coordinates out of the static limit is made.

\end{abstract}

\ccode{PACS numbers: 04.20.-q, 98.80.Jk}

% 04.20.-q Classical general relativity
% 98.80.Jk Mathematical and relativistic aspects of cosmology

%\tableofcontents
%%%% *****************************************************************
\section{Introduction}
\label{secI}

    The existence of  particles with negative energies and the possibility
of observations of consequences of their existence is well known in
the black hole physics due to
the Penrose process.\cite{Penrose69,PenroseFloyd71}
    If one considers the energy of massive particle at rest on
space infinity as equal to $mc^2$ ($c$ is the velocity of light) then
there is an illusion that the negative energy is possible only in case
of the very strong external (for example gravitational) field.
    Really,  for velocity $v \ll c$
the full negative energy $E$ of the particle moving on the distance $r$
from the attracting massive body with the mass $M$
    \begin{equation}    \label{nr}
E = m c^2 + \frac{m v^2}{2} - G \frac{mM}{r} <0 \ \ \Rightarrow \ \
r < \frac{G M}{c^2} = \frac{r_g}{2},
\end{equation}
    where $G$ is gravitational constant, $r_g$ is the gravitational radius.
      However the definition of the energy depends on the choice of
the Killing vector which corresponds to translation in time of
the reference frame.
    This leads to the possibility of existence of negative energy in
noninertial reference frame.

        In papers\cite{GribPavlov2016d,GribPavlov2019} it was shown that
the description of particle movement in rotating coordinates is very
similar to the description of movement in the field of rotating black hole
in Boyer-Lindquist coordinates.\cite{BoyerLindquist67}
    In the rotating coordinates also exists a surface called the static
limit out of which no body can be at rest in the chosen coordinates.
    Out of the static limit in rotating coordinate system the processes
similar to Penrose process in black holes are possible.

    Here we shall discuss the third possibility of existence of
particle states with negative energy --- that of the expanding Universe.

%%%% *****************************************************************
\section{Negative Energies in Expanding Universe}
\label{secNEEU}

    Take the interval of the Friedmann homogeneous and isotropic
expanding Universe in the Robertson-Walker form
    \begin{equation}
d s^2 = c^2 d t^2 - a^2(t)
\left( \frac{d r^2}{1 - k r^2} + r^2 d \Omega^2 \right),
\label{f1}
\end{equation}
    where $k = \pm 1, 0$ for three types of Friedmann models,
$d \Omega^2 = d \theta^2 + \sin^2 \theta \, d \varphi^2 $.
    In coordinates $t, r, \theta, \varphi$ the background matter defining
the metric of space-time is at rest.
%%%%%%%%%%%%%%%%%%%%%%%%%%%%%%%%%%%%%%%
    For simplicity take the case $ k=0$.
    Then the distance from the observer at rest relative to the background
matter at the point $r=0$, for example of the astronomer on
the nonrotating Earth, to the object with the coordinate $r$ will be defined
by the product $ a r$ (see Refs.~\refcite{Ellis93,Grib95}).
    Let us use the new coordinates $t, D, \theta, \varphi$:
    \begin{equation}
D = a(t)\, r, \ \ \ \
d D = \dot{a} r\, d t + a \, d r.
\label{f2}
\end{equation}
    Then
    \begin{equation}
d r = \frac{d D}{a} - \frac{D \dot{a}}{a^2}\, d t
\label{f3}
\end{equation}
    and the interval~(\ref{f1}) becomes
    \begin{equation}
d s^2 = \left( 1 - \frac{D^2 \dot{a}^2}{c^2 a^2 } \right) c^2 d t^2
+ \frac{2 D \dot{a}}{a} d D d t - d D^2 - D^2 d \Omega^2 .
\label{f4}
\end{equation}

    Consider particle movement in metric of the general form
    \begin{equation}
d s^2 = g_{00} (dx^0)^2 + 2 g_{01} dx^0 dx^1 + g_{11} (dx^1)^2 + g_{\Omega \Omega} d \Omega^2,
\label{f5}
\end{equation}
    where $g_{11} < 0 $, $ g_{\Omega \Omega} < 0 $ and $g_{00} g_{11} -g_{01}^2 < 0 $.
    From the condition $d s^2 \ge 0$ for any possible
displacements of particles  one obtains the limitation on
the velocity component
    \begin{eqnarray}
&&\frac{g_{01} - \sqrt{ g_{01}^2 - g_{11} g_{00} - g_{11} g_{\Omega \Omega}
\left( \frac{d \Omega}{ d x^0 } \right)^2 } }{-g_{11}}
\le \frac{d x^1}{d x^0}  \le
\nonumber \\
&&\hspace*{5cm} \frac{g_{01} + \sqrt{ g_{01}^2 - g_{11} g_{00} - g_{11} g_{\Omega \Omega}
\left( \frac{d \Omega}{ d x^0 } \right)^2 } }{-g_{11}} .
\label{f6}
\end{eqnarray}
    One can see from~(\ref{f6}) that the surface  $g_{00}=0$  plays
the role of the static limit.
    In the chosen coordinate system in the region $g_{00} <0$
no physical body can be at rest.
    In this region movement will be observed either from the observer
if  $g_{01} > 0$  corresponding to the expanding Universe~(\ref{f4})
with  $\dot{a}>0$ or to the observer for  $g_{01} < 0$ in case
of contracting Universe~(\ref{f4}) with $\dot{a}<0$.

    The particle energy for the free movement in space-time with
metric~(\ref{f5}) is defined by the canonical 4-momentum  $p_i$
    \begin{equation}
E= p_0 c = mc g_{0k} \frac{d x^k}{ d \tau} =
mc \frac{d x^0}{ d \tau} \left( g_{00} + g_{01} \frac{d x^1}{ d x^0} \right).
\label{f7}
\end{equation}
    Here $\tau $ is the particle proper time.
    If the metric does not depend on time then the energy is conserved.
    The necessary condition for the energy to be negative is
    \begin{equation}
\frac{d x^1}{ d x^0}  < - \frac{g_{00}}{g_{01} }.
\label{f7dop}
\end{equation}
    Using the limitation~(\ref{f6}) one obtains the limitations
on possible values of the energy of particle
    \begin{eqnarray}
\nonumber
\frac{ m c}{-g_{11}} \frac{ d x^0 }{d \tau }
\left( g_{01}^2 - g_{11} g_{00} - |g_{01}| \sqrt{ g_{01}^2 - g_{11} g_{00} -
g_{11} g_{\Omega \Omega} \left( \frac{d \Omega}{ d x^0 } \right)^2 }\right)
\le E \le  \\
\frac{ m c}{-g_{11}} \frac{ d x^0 }{d \tau }
\left( g_{01}^2 - g_{11} g_{00} + |g_{01}| \sqrt{ g_{01}^2 - g_{11} g_{00} -
g_{11} g_{\Omega \Omega} \left( \frac{d \Omega}{ d x^0 } \right)^2 }\right).
\label{f8}
\end{eqnarray}
    From inequalities~(\ref{f8}) one has that at the region
where $g_{00} > 0$ the particle energy is always positive,
on the static limit the minimal energy can be zero,
in the region $g_{00} <0$  (out of the static limit)  states with zero
and negative energy with any absolute value are possible.

    For the metric~(\ref{f4}) the static limit is
    \begin{equation}
D_s = \frac{ c a }{|\dot{a}|} = \frac{c}{|h(t)|},
\label{fps}
\end{equation}
    where $ h(t) = \dot{a}/a $ is the Hubble parameter.
    The energy of the freely moving particle is
    \begin{equation}
E= E' \left( 1 - \frac{\dot{a}^2}{a^2 } \frac{D^2}{c^2} +
\frac{\dot{a}}{a} \frac{D}{c^2} \frac{d D}{d t} \right),
\label{f9}
\end{equation}
    where $E' = m c^2 dt / d \tau $.
    So in these coordinates particles with negative energies
are moving in coordinate $D > D_s$ so that the velocity is slower
than some definite value
    \begin{equation}
\frac{d D}{ d t} < \frac{ c^2 a}{D \dot{a} } \left( \frac{D^2}{D_s^2} - 1 \right).
\label{f10}
\end{equation}
    Let us rewrite the inequality~(\ref{f10}) in terms of coordinates
$t, r, \theta, \varphi$.
    Then we obtain in the case $\dot{a} > 0$
    \begin{equation}
v = a \frac{d r}{d t} < - c \frac{D_s}{D}, \ \ \ D > D_s.
\label{vfon}
\end{equation}
    This has a meaning similar to that obtained by us for
the case of rotating coordinate frame $t, r, \theta, \varphi$:
particles with negative energies close to the static limit  ($D \to D_s$)
must move with velocities close to the light velocity in direction
of the observer in expanding Universe.

    Now let us discuss a special case of the  de Sitter Universe
with the scale factor $a = a_0 \exp H t$,
where $a_0$ and $H$ are constants.
    The Hubble constant is constant and the static limit $D_s = c/H$
is constant and it is equal to the cosmological event horizon for
the observer at the origin
    \begin{equation}
L_H = a(t) c \int \limits_{t}^{\infty} \frac{d t'}{a(t')} = \frac{c}{H} .
\label{aKH}
\end{equation}
    So processes (Penrose processes) with particles with negative energy
are not seen by the observer.
    This is analogous to situation of nonrotating black holes where particles
with negative energy exist only inside the event
horizon.\cite{GribPavlov2010NE}
    However for scale factor $a = a_0 t^\alpha$ ($ 0 < \alpha <1 $)
the cosmological horizon does not exist and some visible consequences of
existence of particles with negative energies can be observed.

    In this paper we consider particles as classical particles.
    Surely if they are quantum new features appear.\cite{Vilenkin80,Neznamov18a}

\section*{Acknowledgments}

This research is supported by the Russian Foundation for Basic research (Grant No. 18-02-00461 a).
The work of Yu.V.P. was supported by the Russian Government Program of Competitive Growth
of Kazan Federal University.

%%%% *****************************************************************

\end{document}